\documentclass[referee]{raa}    
\usepackage{graphicx,times}
\usepackage{natbib}
\usepackage{amssymb,amsmath}
\usepackage{embrac}
\usepackage{indentfirst}
\bibpunct{(}{)}{;}{a}{}{,}

\usepackage{color}

\newcommand\rsun{\mbox{R$_\odot$}}%

\newcommand\kms{\mbox{km s$^{-1}$}}%

\begin{document}

   \title{Analysis of the distribution, rotation and scale characteristics of solar wind switchbacks: comparison between the first and second encounters of Parker Solar Probe

}

 \volnopage{ {\bf 20XX} Vol.\ {\bf X} No. {\bf XX}, 000--000}
   \setcounter{page}{1}

   \author{Mingming Meng\inst{1,2}, Ying D. Liu\inst{1,2}, Chong Chen\inst{1,2}, Rui Wang\inst{1}

   }

   \institute{ State Key Laboratory of Space Weather, National Space Science Center, Chinese Academy of Sciences, Beijing 100190,
People's Republic of China; {\it liuxying@swl.ac.cn}
\\
        \and
             University of Chinese Academy of Sciences, No.19A Yuquan Road, Beijing 100049, People's Republic of China\\
   }
\vs \no

\abstract{The S-shaped magnetic structure in the solar wind formed by the twisting of magnetic field lines is called a switchback, whose main characteristics are the reversal of the magnetic field and the significant increase in the solar wind radial velocity. We identify 242 switchbacks during the first two encounters of Parker Solar Probe (PSP). Statistics methods are applied to analyze the distribution and the rotation angle and direction of the magnetic field rotation of the switchbacks. The diameter of switchbacks is estimated with a minimum variance analysis (MVA) method based on the assumption of a cylindrical magnetic tube. We also make a comparison between switchbacks from inside and the boundary of coronal holes. The main conclusions are as follows: (1) the rotation angles of switchbacks observed during the first encounter seem larger than those of the switchbacks observed during the second encounter in general; (2) the tangential component of the velocity inside the switchbacks tends to be more positive (westward) than in the ambient solar wind; (3) switchbacks are more likely to rotate clockwise than anticlockwise, and the number of switchbacks with clockwise rotation is 1.48 and 2.65 times of those with anticlockwise rotation during the first and second encounters, respectively; (4) the diameter of switchbacks is about $10^{5}$ km on average and across five orders of magnitude ($10^{3}$ -- $10^{7}$ km).
}+
   \authorrunning{Meng et al.}            
   \titlerunning{Analysis of the distribution, rotation and scale of switchbacks}  
   \maketitle

\section{Introduction}           
\label{sect:intro}

The Parker Solar Probe (PSP) mission launched in 2018 is the closest spacecraft to the Sun in history \citep{2016Fox}. During the first two encounters of PSP, a large number of sudden radial magnetic field reversals associated with velocity spikes, namely switchbacks, are detected \citep{2019Bale,2019Kasper}. This is not the first observation of switchbacks. Switchbacks have been observed by Ulysses in 1994 -- 1995 (e.g., \citealt{1996Forsyth,1999Balogh,2002Yamauchi}) and Helios (e.g., \citealt{2018Horbury,2020Macneil}). However, PSP observes that switchbacks are ubiquitous and are in the early stage of evolution at a higher time resolution in the solar wind, which enables us to make more detailed analysis about the origin, characteristics and evolution of switchbacks.

Switchbacks exist widely in the solar wind and are obviously different from other forms of the change of the magnetic field polarity, such as crossings through a magnetic loop. Based on the hourly averaged magnetic field data from Ulysses during 1994 -- 1995, \citet{1999Balogh} find that about 8.4\% of the observation time the magnetic fields deviate from Parker spiral more than 90°. \citet{2003Yamauchi} suggest that, when spacecraft passes through different types of magnetic field reversals, different pitch angle distributions of suprathermal electrons will be observed, which could help distinguish between switchbacks and other types of magnetic field reversals. Another characteristic of switchbacks is that, when spacecraft observes a switchback, it could also observe that the speed of protons is higher than the speed of alpha particles (e.g., \citealt{2004Yamauchia,2013Neugebauer}).

Switchbacks are generally highly Alfvénic, so they are suggested as outward-propagating Alfvénic structures (e.g., \citealt{2019Bale,2019Kasper,2020Horbury}). \citet{2020Dudok} suggest that switchbacks originate from the corona, which is supported by some simulation results (e.g., \citealt{2021Drake,2020Ruffolo,2020Fisk}). Interchange magnetic reconnection is one of the probable trigger mechanisms of switchbacks (e.g., \citealt{2004Yamauchib,2021Drake,2020Fisk}). In addition, shear-driven turbulence around the Alfvén critical zone \citep{2020Ruffolo} and footpoint motion of the solar wind between fast and slow streams \citep{2021Schwadron} could product switchbacks. These also support switchbacks' coronal origin hypothesis. \citet{2020Tenerani} show that, if the background is homogeneous, switchbacks originating in the corona could propagate out to PSP. Switchbacks could also be triggered in situ in the solar wind. \citet{2020Squire} suggest that low-amplitude outward-propagating waves could naturally develop into switchbacks in the process of spreading out. \citet{2020Mozer} suggest that the number and the rotation angle of switchbacks would be more and larger with the increase of heliocentric distances base on PSP observation on 2019 April 5 and 2019 March 31. A similar trend in the number of switchbacks is also observed by Helios \citep{2020Macneil}. This trend indicates that these switchbacks may form in situ as the solar wind spreads out.

Switchbacks triggered by different physical processes may have differences. In this study, we make a detailed analysis of the distribution, rotation and scale of switchbacks observed by PSP during the first two encounters, and compare the differences in the switchbacks between the first and second encounters. This paper is organized as follows. Section \ref{sect:Data} shows the observation data and the methods. Section \ref{sect:Analysis} gives the results from our analysis of switchbacks' distribution, rotation and scale characteristics. We discuss the results and conclude in Section \ref{sect:discussions}. Our results may help understand the origin and properties of solar wind switchbacks.

\section{Data and methodology}
\label{sect:Data}

The magnetic field is measured by the fluxgate magnetometer (MAG) from the FIELDS suite \citep{2016Bale}, and the plasma parameters are measured by the Solar Probe Cup (SPC) from the SWEAP package \citep{2016Kasper}. The perihelion of the first and second encounters is at a distance of 35.7 \rsun{} from the center of the Sun, and the data that we use cover 35.7 -- 56.1 \rsun{} and 35.7 -- 55.1 \rsun{} for the first and second encounters respectively. Both the magnetic field and plasma data are from  the first two encounters and are  in the  RTN frame. The magnetic field data are interpolated to the plasma resolution. The photospheric field synoptic maps used in this work are provided by the Global Oscillation Network Group (GONG). The coronal  193 \AA{} images are from the  Atmospheric Imaging Assembly (AIA;  \citealt{2012Lemen}) on board the Solar Dynamics Observatory (SDO; \citealt{2012Pesnell}).

The criteria used to identify switchbacks are as follows: (1) magnetic field has a rotation relative to the ambient solar wind; (2) the radial velocity shows an enhancement; and (3) the pitch angle distribution (PAD) of the 315 eV suprathermal electrons is unidirectional across a switchback. We first screen magnetic field intervals that the sign of radial magnetic field changes. Then, we examine intervals based on the magnetic field, velocity and the electron PAD characteristics of typical switchbacks. Both the magnetic field rotation and the enhancement in the radial velocity are relative to the ambient solar wind. Intervals, whose data points are less than five, are removed. We identify 129 switchbacks during the first encounter and 113 switchbacks during the second encounter. It is worth noting that the minimum time interval  of these switchbacks is about 1.966 seconds. Therefore, the change of resolution during the encounters  would not affect our results.

A potential field source surface  (PFSS) Model \citep{1969Altschuler,1969Schatten,1984Hoeksema,1992Wang} and a  Parker spiral model \citep{1958Parker} are used to trace the footpoints  of switchbacks. The  PFSS model is included in the Solar Software (SSW) \citep{1998Freeland}. We trace the field footpoints using  equation \eqref{eq1}. Here  $\Phi$ and $\Phi_{0}$ are the Carrington longitude of the photosphere footpoint and PSP, respectively, $\Omega$ is the angular speed of the Sun, L is the distance from PSP to the Sun, and V is the speed of the solar wind from PSP measurements. In this paper we adopt a traceability method similar to \citet{2018Zhu} and \citet{2020Badman}, which first traces the Carrington coordinates of switchbacks from the location of PSP to 2.5 \rsun{} based on the Parker Spiral model and then determines the footpoint of switchbacks with the PFSS model.

\begin{equation} 
        \label{eq1}
        \Phi=\Phi_{0}+\Omega{}L/V
\end{equation}

\begin{figure}[!htbp]
    \centering
    \includegraphics[width=0.60\textwidth]{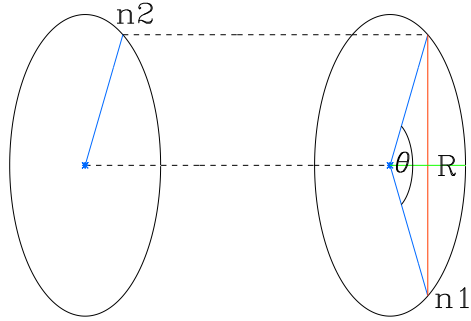}
    \caption{Sketch showing the estimate of the diameter of switchbacks. Here n1 and n2 represent the normal direction at the positions that PSP enters and leaves a switchback respectively. The normal direction points to the center along the blue line. R is the distance perpendicular to the axial direction of the cylinder that PSP passes through.}
    \label{fig:1}
\end{figure}

\citet{2020Horbury} suggest that a switchback corresponds to a long and thin cigar-like structure. \citet{2021Laker} assume a cylindrically symmetric, long, thin structure to analyze the width and aspect ratio of switchbacks. Based on this assumption, we analyze the diameter of switchbacks using a minimum variance analysis (MVA). The details are as follows. As Figure \ref{fig:1} shows, if we conceive that a switchback is a cylinder, the distance that PSP passes through a switchback can be projected into two directions that are parallel to the axis of the cylinder, and perpendicular to the axis of the cylinder. The distance perpendicular to the axial direction of the cylinder provides the information about the diameter of the cylinder. However, we do not know the distance from PSP's trajectory to the cylinder axis. The MVA is used to determine the normal direction of an interval \citep{1967Sonnerup}, which gives us a way to estimate the diameter of the cylinder based on the distance perpendicular to the axial direction of the cylinder that PSP crosses. The boundary of switchbacks corresponds to the cylinder surface. On the cylinder surface the normal direction points to the axis of the cylinder. The axial direction of the cylinder is approximately represented by the direction of the average magnetic field in switchbacks. We can obtain the normal directions n1 and n2 by MVA on the boundaries of switchbacks, and translate these two normal directions into the same plane. Here $\theta{}$ is the angle between n1 and n2. The velocity of switchbacks can be projected into the directions parallel and perpendicular to the axial direction. R can be calculated by the velocity that is perpendicular to the axial direction. In order to avoid the influence of the data fluctuations on the judgment of the normal direction, filtering is applied to the boundary. We use equation \eqref{eq2} to determine the quality of the diameter measurement. The diameter of a switchback is given by equation \eqref{eq3}.

\begin{equation}
        \label{eq2}
        \Delta{}\varphi{}_{ij}=\sqrt{\lambda_{3}(\lambda_{i}+\lambda_{j}-\lambda_{3})/(M-1)( \lambda_{i}-\lambda_{j})^{2}}
\end{equation}

Generally, if $\Delta{}\varphi{}_{ij}$ is less than 0.087, the result of MVA is considered to be reliable \citep{2012chou}. Here $\lambda$ is the eigenvalue of the covariant matrix, i and j represent Cartesian field components and i $\neq$ j, and M represents the number of data points used for MVA.

\begin{equation}
        \label{eq3}
        D=\frac{R}{sin\frac{\theta{}}{2}}
\end{equation}

\section{Analysis and results}
\label{sect:Analysis}

\begin{figure}[!htbp]
    \centering
    \includegraphics[width=1.00\textwidth]{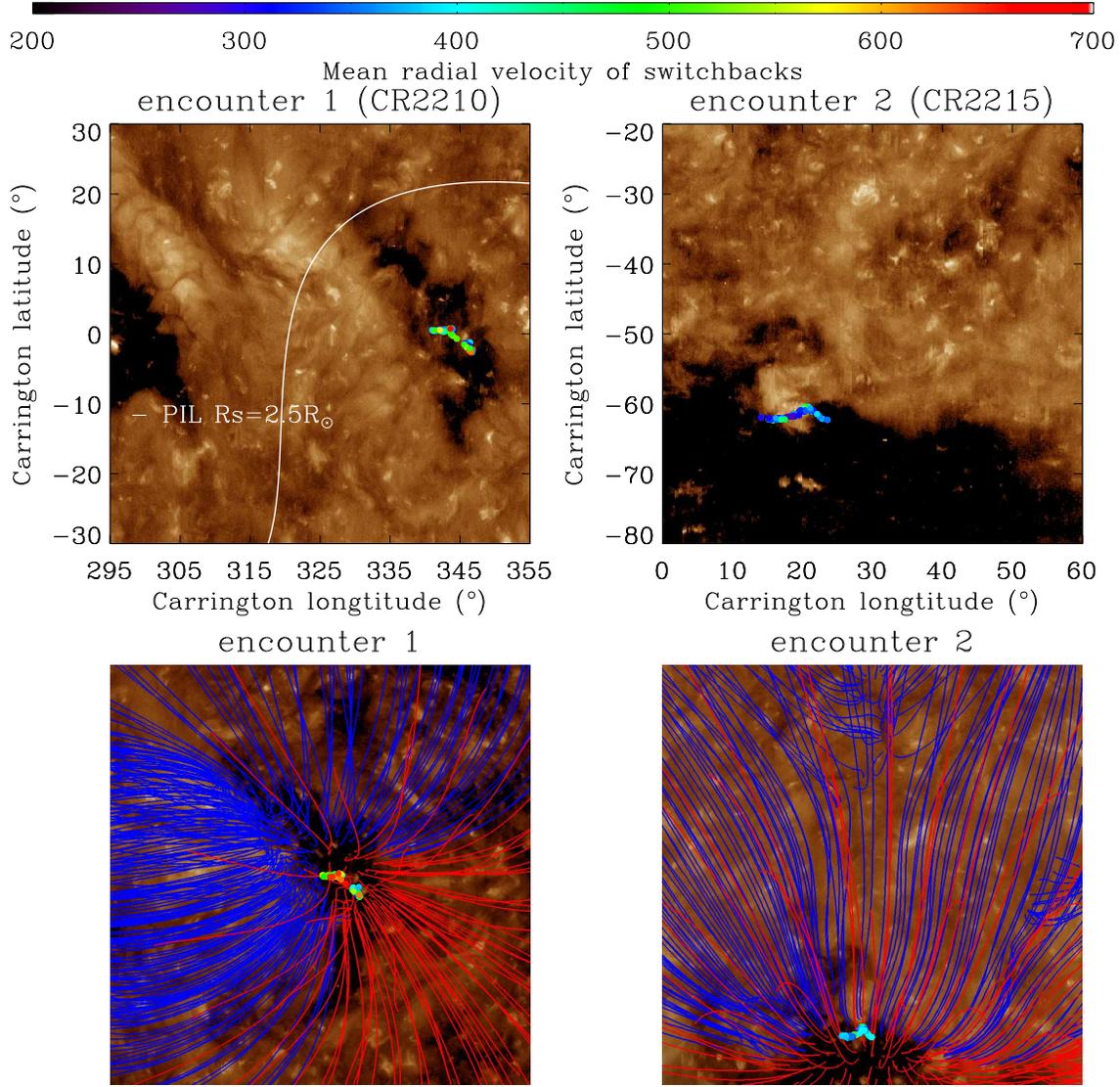}
    \caption{Footpoints of switchbacks observed during the first (left) and the second (right) encounters projected on the AIA 193 \AA{} images. Upper panels show footpoints of switchbacks only and lower panels add coronal magnetic field lines. The white line in the upper left panel is the neutral line. Points with different colors represent footpoints of switchbacks, and the colors represent the mean radial velocity of switchbacks. Blue lines represent closed magnetic field lines. Red lines are inward open magnetic field lines.}
    \label{fig:2}
\end{figure}

Figure \ref{fig:2} shows the footpoints of the switchbacks observed during the first and second encounters. The source regions on the solar disk of switchbacks observed during the first and second encounters are different. As shown in Figure \ref{fig:2}, switchbacks observed during the first encounter mainly origin from a small coronal hole, which is consistent with \citet{2019Bale} and \citet{2020Badman}. The footpoints of switchbacks observed during the second encounter are from the interface between closed and open magnetic field lines. The locations of the footpoints are also consistent with AIA 193 \AA{} images. Interchange reconnection is likely to occur between the open and closed field lines. \citet{2020Rouillard} show that PSP observed streamer flows during the second encounter.

\begin{figure}[!htbp]
    \centering
    \includegraphics[width=1.00\textwidth]{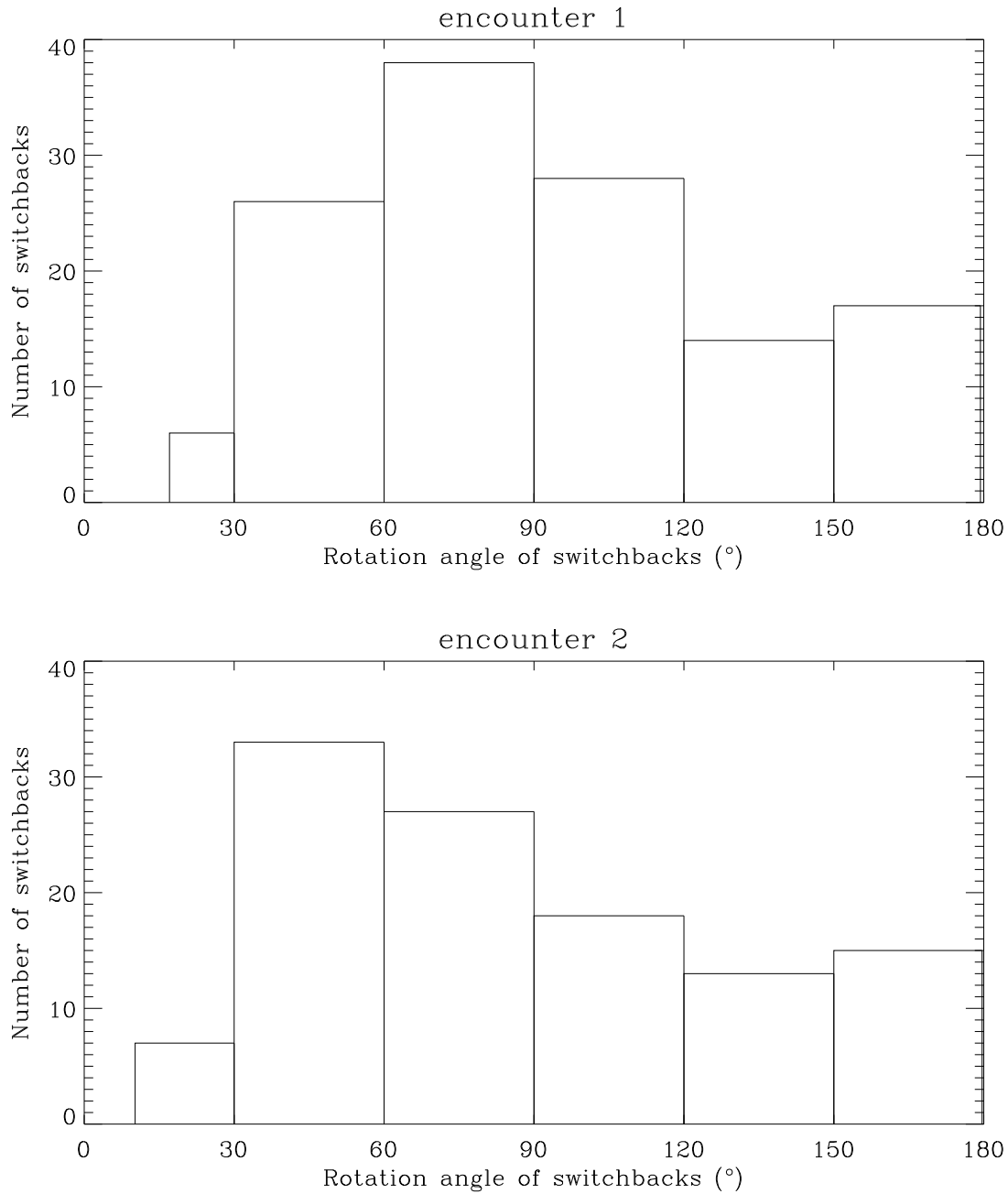}
    \caption{The distribution of switchbacks in terms of rotation angles for encounter 1 (upper) and 2 (lower).}
    \label{fig:3}
\end{figure}

We make a detailed analysis of rotation angles of switchbacks. Figure \ref{fig:3} shows the distribution of rotation angles. From the top panel of Figure \ref{fig:3}, we can see that the number of switchbacks with a rotation angle (60 -- 90°) seems to be the most during the first encounter. The number of switchbacks with a small rotation angle (0 -- 60°) is almost equal to the number of switchbacks with a large rotation angle (120 -- 180°). From the bottom panel of Figure \ref{fig:3}, we can see that the rotation angles of switchbacks observed during the second encounter seem smaller than those of the switchbacks observed during the first encounter in general. The rotation angles of switchbacks are mainly concentrated at 30 -- 90° during the second encounter.

\begin{figure}[!htbp]
    \centering
    \includegraphics[width=1.00\textwidth]{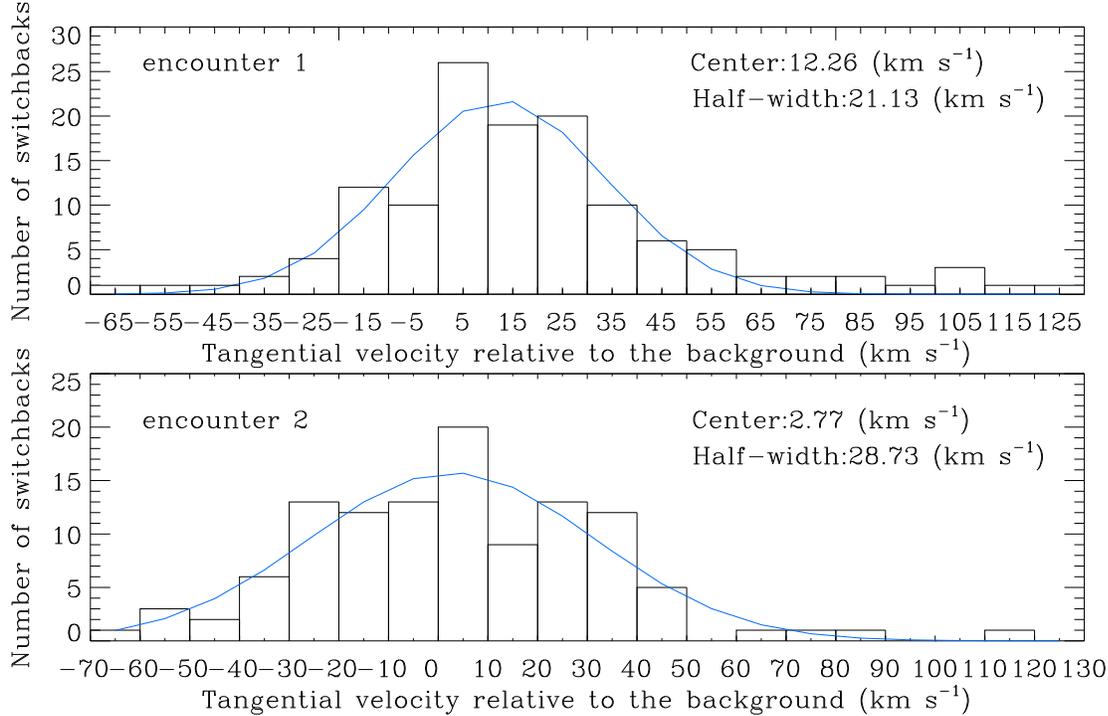}
    \caption{Distribution of the tangential velocity component within switchbacks relative to the background solar wind during the first and the second encounters. The blue curve shows a Gaussian fit.}
    \label{fig:4}
\end{figure}

We find that the tangential component of the velocity may also have a tendency. Figure \ref{fig:4} shows the distribution of the tangential velocity component inside switchbacks relative to the background solar wind during the first and second encounters respectively. From the figure we can see that the tangential component of the velocity inside the switchbacks tends to be more positive. The center and half-width of a Gaussian fit of the first (second) encounter are 12.26 (2.77) and 21.13 (28.73) \kms{} respectively. In the first encounter, some switchbacks have a significant variation in the velocity tangential component, up to 100 \kms{}. The reason for this tendency of being more positive is still unclear. We think that this tendency may be related to the direction of rotation of switchbacks.

We then make a statistical analysis on the rotation direction of switchbacks. The rotation angle relative to the ambient solar wind is limited to 0 to 180° and the sign of the rotation angle represents the direction of rotation. We focus on the rotation of switchbacks in the RT plane. The positive rotation angle corresponds to anticlockwise rotation and the negative rotation angle corresponds to clockwise rotation. Statistical results are shown in the Table \ref{tab:1}. We can see that switchbacks are more likely to rotate clockwise during the first two encounters, which is similar to what is observed by Helios \citep{2020Macneil}. The results show that 77 switchbacks rotate clockwise and 52 switchbacks rotate anticlockwise during the first encounter, and 82 switchbacks rotate clockwise and 31 switchbacks rotate anticlockwise during the second encounter. The number of switchbacks with clockwise rotation is 1.48 (2.65) times of those with anticlockwise rotation during the first (second) encounter. The direction of rotation of switchbacks is directly influenced by the trigger mechanism of switchbacks. Some trigger mechanisms could just produce clockwise rotations, such as velocity shear or magnetic field line draping by ejecta (see Figure 3 of \citealt{2020Macneil}).

\begin{table}[!htbp]
    \caption{Distribution of rotation direction of switchbacks.}
    \label{tab:1}
    \centering
    \footnotesize
    \setlength{\tabcolsep}{4pt}
    \renewcommand{\arraystretch}{1.2}
    \begin{tabular}{lcccccccc}

\hline
Statistics of the direction of switchbacks       &Number of switchbacks\\
\hline
clockwise rotation (encounter 1)	     &77\\
anticlockwise rotation (encounter 1)	 &52\\
clockwise rotation (encounter 2)	     &82\\
anticlockwise rotation (encounter 2)	 &31\\
        \hline
    \end{tabular}
\end{table}

\begin{figure}[!htbp]
    \centering
    \includegraphics[width=1.00\textwidth]{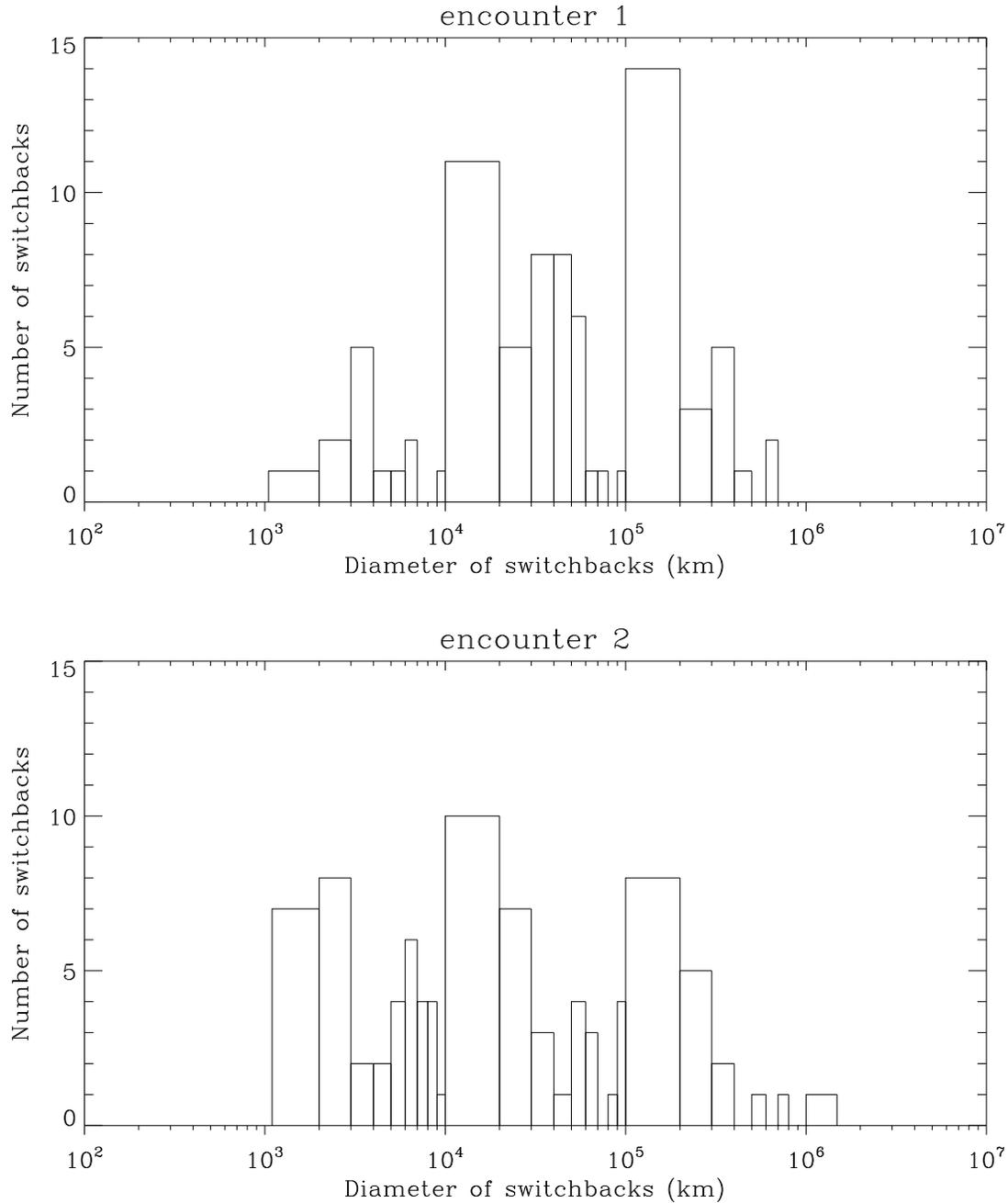}
    \caption{Distribution of diameters of switchbacks measured during the first (upper) and second (lower) encounters.}
    \label{fig:5}
\end{figure}

Using the method shown in Figure \ref{fig:1} we analyze the diameters of switchbacks. Just the switchbacks whose $\Delta{}\varphi{}_{ij}$ is less than 0.087 are kept. The results are shown in Figure \ref{fig:5}. Seventy-nine and eighty-nine switchbacks are reserved, respectively, during the first and second encounters. During the first encounter, the minimum diameter of switchbacks is about 1,000 km, the maximum diameter of switchbacks is about 640,000 km, and the average diameter of switchbacks is about 100,000 km. During the second encounter, the minimum diameter of switchbacks is about 1,000 km, the maximum diameter of switchbacks is about 1,500,000 km, and the average diameter of switchbacks is about 80,000 km. The diameter of switchbacks spans five orders of magnitude from $10^3$ to $10^7$ km. These are consistent with the results of \citet{2021Laker} and \citet{2021Larosa}. It is not ruled out that smaller switchbacks may not have been observed. In general, the diameters of most switchbacks are between $10^{4}$ -- $10^{6}$ km from inside the small coronal hole during the first encounter and $10^{3}$ -- $10^{5}$ km from the boundary of the large coronal hole during the second encounter.

\section{Discussions and conclusions}
\label{sect:discussions}

PSP approximately corotates with the Sun during the first encounter and mainly observes solar wind streams from a small coronal hole. During the second encounter PSP mainly observes streamer flows. This gives us a great opportunity to compare the differences of switchbacks from inside and the boundary of coronal holes. In this paper, we make a statistical analysis of the distribution, rotation angle and direction, and scale of switchbacks observed during the first two encounters. The focus of our statistical analysis is to compare the differences between switchbacks from inside and the boundary of the coronal holes. The main results are as follows.

PSP mainly observed solar wind streams from a small coronal hole during the first encounter, and the footpoints of the switchbacks observed during the first encounter are mainly located inside the small coronal hole. PSP observed streamer flows during the second encounter, and the footpoints of the switchbacks are located between closed and open magnetic field lines. The different solar wind origins may give rise to different properties of switchbacks.

The distribution of rotation angles of switchbacks seems to have differences between the first and second encounters, which may be related to the different origins of switchbacks. The switchbacks observed during the first encounter seem to have larger rotation angles than the switchbacks observed during the second encounter in general. In the first encounter, switchbacks with a rotation angle (60 -- 90°) seems to be the most, and about 45\% of switchbacks have a rotation angle greater than 90°. In the second encounter, switchbacks with a rotation angle (30 -- 60°) seems to be the most, and about 40\% of switchbacks have a rotation angle greater than 90° during the second encounter.

PSP observed large tangential velocities that exceed the value from the axisymmetric Weber-Davis model \citep{2019Kasper}. Our results indicate that the tangential component of the velocity in the switchbacks tends to be more positive than in the ambient solar wind. The cause for this tendency is still unclear. It is an interesting issue if there is any relation between switchbacks and the tangential flow.

Switchbacks have a tendency in the direction of rotation. Switchbacks generally tend to rotate clockwise from both inside and the boundary of the coronal holes, which is also similar to the observation made by Helios \citep{2020Macneil}. This trend reflects the different trigger mechanisms of switchbacks. \citet{2020Macneil} introduce the relationship between the trigger mechanism and the direction of rotation, and \citet{2021Larosa} give a detailed summary about the different trigger mechanisms of switchbacks. The physical process that could only produce switchbacks with clockwise rotation, such as local velocity shear, may be an important trigger mechanism, because the observed switchbacks are more likely to rotate clockwise. The direction of rotation and the change of plasma parameters of switchbacks could help make a detailed distinction between the different trigger mechanisms of switchbacks.

We also make a statistical analysis of the diameter of switchbacks. The diameter of switchbacks is about $10^5$ km on average, which is similar to the results of \citet{2021Laker}, and the magnitude of the diameter crosses five orders ($10^{3}$ -- $10^{7}$ km). In addition, \citet{2021Laker} suggest that the mean length of switchbacks is about 500,000 km, corresponding to an aspect ratio of the order of 10. The diameter of switchbacks seems larger from inside a coronal hole than from the boundary of a coronal hole in general.

\normalem
\begin{acknowledgements}
The research was supported by NSFC under grants 41774179 and 12073032, Beijing Municipal Science and Technology Commission (Z191100004319003), and the Specialized Research Fund for State Key Laboratories of China. We acknowledge the data provided by PSP, SDO, and GONG. We also thank Bei Zhu, Lei Lei, Xiaowei Zhao and Huidong Hu for their help on this work.
\end{acknowledgements}
  
\bibliographystyle{raa}
\bibliography{bibtex}

\end{document}